\documentclass[aps,prl,showpacs,twocolumn,superscriptaddress]{revtex4}
\usepackage{bm}
\usepackage[dvipdfmx]{graphicx}
\usepackage[dvipdfmx]{color}
\usepackage{amsmath, amssymb}
\usepackage{braket}
\usepackage{color}
\usepackage{comment}
\begin{document}
\title {Theoretically proposed controlled creation of Bloch-type skyrmions with spin-orbit torque in a chiral-ferromagnet/heavy-metal heterojunction}
\author{Yuto Uwabo}
\affiliation{Department of Applied Physics, Waseda University, Okubo, Shinjuku-ku, Tokyo 169-8555, Japan}
\author{Masahito Mochizuki}
\affiliation{Department of Applied Physics, Waseda University, Okubo, Shinjuku-ku, Tokyo 169-8555, Japan}
\begin{abstract}
The creation and manipulation of magnetic skyrmions in magnetic bilayer heterostructures via spin-orbit torque have been intensively studied in spintronics because of their potential application as information carriers in next-generation magnetic memory devices. However, experimental attempts have not always been successful. In this paper, we theoretically elucidate the underlying reasons for these difficulties and propose a practical method to overcome them by employing magnetic bilayer heterostructures that incorporate a chiral ferromagnetic layer hosting Bloch-type skyrmions instead of the conventional ferromagnetic layer that hosts N\'{e}el-type skyrmions. Our micromagnetic simulations demonstrate that Bloch-type skyrmions can be controllably created in this system via spin-orbit torque exerted by a perpendicular spin current. This finding provides a promising platform and method for realizing skyrmion-based spintronic devices.
\end{abstract}
\maketitle

\section{Introduction}
\begin{figure}[tb]
\includegraphics[scale=1.0]{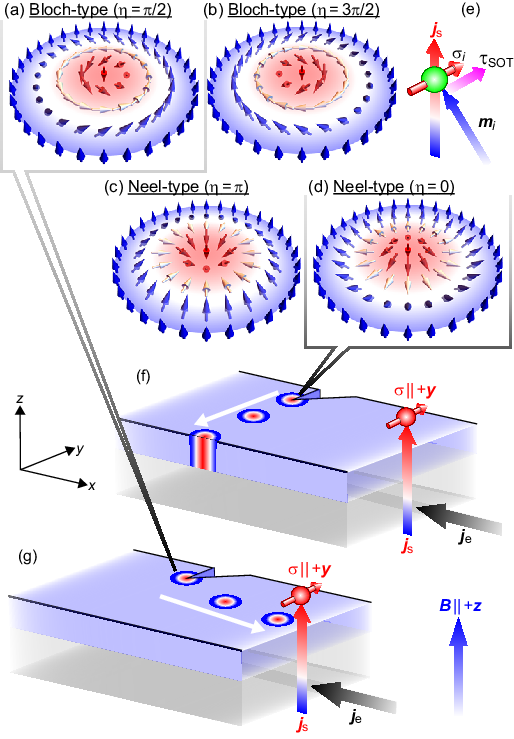}
\caption{(a),~(b) Magnetization configurations of Bloch-type skyrmions with different helicity $\eta$, i.e., (a) $\eta$=$\pi/2$ and (b) $\eta$=$3\pi/2$. (c),~(d) Magnetization configurations of N\'{e}el-type skyrmions with different helicity $\eta$, i.e., (c) $\eta$=$\pi$ and (d) $\eta$=0. (e) Schematic illustration of the spin-orbit torque $\bm \tau_{\rm SOT}$ exerted on local magnetization $\bm m_i$ from a vertical spin current $\bm j_{\rm s}$ with spin polarization $\bm \sigma_i$. (f) Current-induced motion  of N\'{e}el-type skyrmions driven by spin-orbit torque after being created at the notch. They move pependicular to the injected electric current $\bm j_{\rm e}$ and thus in the width direction of the nanotrack, resulting in collison to the longitudinal edge and annihilation. (f) Current-induced motion  of Bloch-type skyrmions driven by spin-orbit torque after creation at the notch. They move nearly parallel to $\bm j_{\rm e}$ and thus along the length direction, resulting in smooth traveling in the nanotrack.}
\label{Fig01}
\end{figure}
Magnetic skyrmions are particle-like topological spin textures~\cite{Nagaosa2013,SekiBook2016} that have been discovered in certain types of ferromagnets~\cite{Muhlbauer2009,YuXZ2010,Everschor2018,Tokura2020}. Since their discovery, they have attracted significant research interest in the field of spintronics~\cite{Fert2013,Koshibae2015,Finocchio2016,KangW2016,Fert2017,XichaoZ2020} due to their topological stability and efficient current-driven mobility. These features make them promising candidates for next-generation memory technologies such as racetrack memory~\cite{Tomasello2014,ZhangX2015,YuG2017,LaiP2017,Maccariello2018,ZhuD2018,HeB2023}. To this end, the current-driven motion of skyrmions~\cite{Fert2013,SeidelBook2016,Ohki2024,Everschor2012,Schulz2012,Jonietz2010,YuXZ2012,Iwasaki2013a,Iwasaki2013b,Sampaio2013,Iwasaki2014a,WooS2016,ZhangX2017,ZhangX2017b,Litzius2020,Reichhardt2022,ZhangX2023a,LuoJ2023,Pham2024} and the skyrmion Hall effect~\cite{Iwasaki2013a,Iwasaki2013b,YangS2024,ZangJ2011,Everschor-Sitte2014,JiangW2017,Litzius2017} have been extensively studied.

Skyrmions emerge in magnetic systems through a competition between ferromagnetic exchange interactions and the Dzyaloshinskii-Moriya (DM) interactions~\cite{Dzyaloshinskii1957,Moriya1960a,Moriya1960b,Fert1980}, typically under an external magnetic field~\cite{Bogdanov1989,Bogdanov1994}. In addition to chiral ferromagnets such as MnSi and Cu$_2$OSeO$_3$~\cite{Muhlbauer2009,YuXZ2010,Munzer2010,YuXZ2011,Seki2012a,Seki2012b,Tokunaga2015,Karube2017}, which host Bloch-type skyrmions characterized by a vortex-like magnetization configuration [Figs.~\ref{Fig01}(a) and (b)], magnetic bilayer heterostructures composed of a ferromagnetic layer and nonmagnetic heavy-metal layer with strong spin-orbit coupling represent another important class of skyrmion-hosting systems~\cite{ChenG2015,Boulle2016,Moreau-Luchaire2016,Soumyanarayanan2017}. In these systems, the DM interactions arise from the broken spatial inversion symmetry at the interface between the two layers, which stabilizes N\'{e}el-type skyrmions with a fountain-like magnetization configuration [Figs.~\ref{Fig01}(c) and (d)]~\cite{Fert2013}.

Spin-transfer torque is a primary physical mechanism for manipulating magnetic skyrmions in monolayer thin-plate and nanotrack samples of bulk chiral magnet~\cite{Ohki2024}, where the spin angular momenta of electrons of spin-polarized electric current are transferred to the noncollinear magnetizations forming the magnetic texture~\cite{Slonczewski1996,ZhangLi2004,Tatara2004}. In contrast, for the manipulation of magnetic skyrmions with electric currents in magnetic heterojunction systems~\cite{Ohki2024}, spin-orbit torque is a main driving force [Fig.~\ref{Fig01}(e)]~\cite{Manchon2019,ShaoQ2021,KimKW2024}. Specifically, when an electric current is injected into the nonmagnetic heavy-metal layer, a pure spin current that flows in the out-of-plane direction is generated via the spin Hall effect induced by the strong spin-orbit interaction. The spin polarization of this spin current is generally aligned parallel to the interface and perpendicular to the injected charge current, which exerts a torque on the magnetizations that constitute the magnetic texture~\cite{Miron2011,RyuKS2013,Emori2013}.

Regarding the creation of magnetic skyrmions using electric currents, methods based on spin-transfer torque have been extensively studied~\cite{Iwasaki2013b,Sampaio2013,TchoeY2012,Yuan2016,YinG2016,Everschor-Sitte2017,Hrabec2017,YuXZ2017,YuXZ2020,Fujimoto2021,Fujimoto2022,WangW2022}. In general, the controlled creation of topological magnetic textures like skyrmions is difficult. This is because skyrmions belong to a topological class distinct from that of the ferromagnetic state so that it is not possible to create them in a ferromagnetic sample through a continuous variation of the magnetization alignment. Instead, it requires a discontinuous change in the spatial magnetization profile, namely, an abrupt flop of the local magnetization. Because this local magnetization reversal involves a high energy cost, the creation of skyrmions is generally difficult. 

However, a theoretical work in Ref.~\cite{Iwasaki2013b} proposed that injecting an electric current into a monolayer nanotrack of chiral magnet with a small notch allows for controlled creation of skyrmions. In fact, the introduction of a notch structure in the nanotrack can mitigate the above topological constraint. Specifically, the inevitable discontinuity in the spatial magnetization profile at the notch enables a gradual rotation of magnetizations rather than an abrupt flop, leading to the desired local magnetization reversal. In this way, skyrmions can be created with a relatively low energy cost or a small electric current density. Another advantage of using a notched nanotrack is the ability to design and control the location of skyrmion creation. This ingenious approach has been theoretically explored not only for the skyrmion creation with electric currents, but also for creation with  magnetic fields~\cite{Mochizuki2017}, electric fields~\cite{Mochizuki2015a,Mochizuki2015b}, and microwave irradiation~\cite{Miyake2020}.

Although the skyrmion creation with electric currents via the spin-transfer torque has been extensively investigated, there are relatively few studies on that via the spin-orbit torque in magnetic heterojunction systems~\cite{Buttner2017,LiuJ2022}. This difference is attributable to the different behaviors of skyrmions after their creation between the two cases. When skyrmions are created at the corner of a notch in a nanotrack sample via the spin-transfer torque, they move along the longitudinal direction of the nanotrack, following the flow of electrons, thereby completing the creation process in a straightforward manner. In contrast, the situation is more complex when using spin-orbit torque in notched magnetic heterojunction systems. In experiments in Ref.~\cite{Buttner2017}, a larger electric current pulse is first applied to the heterojunction to create a skyrmion, followed by a smaller current to drive its motion. The use of two different current magnitudes is necessary because N\'{e}el-type skyrmions that emerge in heterojunction systems move perpendicular to the injected electric current, that is, in the width direction of the nanotrack rather than its length direction [Fig.~\ref{Fig01}(f)]. Since a relatively strong current is required to create skyrmions, continued injection of this current would drive the skyrmions toward the longitudinal edge of the track, where they may be annihilated through collision and absorption. Therefore, a rather complex procedure and precise control of the electric current are required for skyrmion creation via spin-orbit torque. Specifically, two different current amplitudes must be employed, and the duration of each current pulse must be carefully adjusted to ensure both the successful creation of skyrmions and their stability after creation.


\begin{figure}[tb]
\includegraphics[scale=1.0]{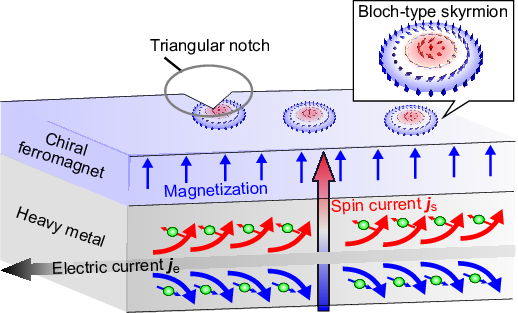}
\caption{Proposed chiral-ferromagnet/heavy-metal bilayer heterojunction with a triangular notch. The chiral-ferromagnet layer can host Bloch-type skyrmions. Conversion of the injected horizontal electric current to the vertical spin current via the spin Hall effect is also shown schematically.}
\label{Fig02}
\end{figure}
In this paper, we propose an alternative and practical method for controlled creation of skyrmions via current injection into a bilayer heterojunction system utilizing the spin-orbit torque mechanism. Considering that the primary difficulty arises from the fact that N\'{e}el-type skyrmions move perpendicular to injected electric current when driven by spin-orbit torque [Fig.~\ref{Fig01}(f)], we examine a method using  Bloch-type skyrmions. Because Bloch-type skyrmions driven by spin-orbit torque  move nearly parallel to the electric current [Fig.~\ref{Fig01}(g)], it is expected that they can smoothly travel in the track after creation. We employ a magnetic heterojunction nanotrack with a triangular notch composed of a chiral ferromagnet hosting Bloch-type skyrmions placed on top of a nonmagnetic heavy-metal layer as illustrated in Fig.~\ref{Fig02}. Micromagnetic simulations demonstrate that by simple injection of electric current into this nanotrack system, we can achieve the controlled creation of Bloch-type skyrmions via spin-orbit torque without any fine-tuning of current-pulse parameters. Our finding offers a highly controllable skyrmion-host system, which provides a promising platform for future skyrmion-based spintronic devices.

\section{Model and Simulation}
\begin{table}
\caption{Unit conversions when $J$=1 meV and $|\bm S_i|/\hbar$=1 (i.e., $M_{\rm s}=\gamma\hbar/a^3$).}
\centering
\begin{tabular*}{8.6cm}{lcc}
\hline \hline
& Dimensionless & SI units \\
& quantity &  \\
\hline
Time                & $\tau$=1000  & $t=10^3M_{\rm s}a^3/\gamma J$=0.66 ns\\
Magnetic field & $\bar{B}_z/J$=0.01  & $B_z=10^{-2}J/M_{\rm s}a^3$=86.4 mT \\
\hline \hline
\end{tabular*}
\end{table}
To depict the chiral-ferromagnet layer of the magnetic bilayer heterojunction, which hosts Bloch-type skyrmions, we employ a classical Heisenberg model on a cubic lattice~\cite{YiSD2009,Mochizuki2012,Buhrandt2013,Tanaka2020}. The Hamiltonian comprises the ferromagnetic exchange interactions, the Zeeman interactions, and the DM interactions as,
\begin{align}
\mathcal{H}&=
-J\sum_{\langle i,j\rangle} \bm m_i \cdot \bm m_j
-\bar{\bm B}_{\rm ext} \cdot \sum_i \bm m_i
\nonumber \\
&-D\sum_i  \left[
  \hat{\bm x} \cdot (\bm m_i \times \bm m_{i+ \hat{\bm x}})
+\hat{\bm y} \cdot (\bm m_i \times \bm m_{i+ \hat{\bm y}}) \right],
\label{eq:Hamilt}
\end{align}
where $\bm m_i$ is the normalized classical magnetization vector at site $i$ ($|\bm m_i|$=1), and $\langle i,j\rangle$ denotes adjacent site pairs. Both the ferromagnetic exchange and DM interactions work on the nearest-neighbor bonds connecting adjacent sites, while the external magnetic field $\bar{\bm B}_{\rm ext}=(0,0,\bar{B}_z)$ is applied perpendicular to the layer, which stabilize Bloch-type skyrmions with helicity of $+\pi/2$. The skyrmion spacing $\lambda_{\rm m}$ in the crystalized phase is determined by the ratio $D/J$. We adopt $D/J$=0.27 in this study, for which the skyrmion spacing $\lambda_{\rm m}$ is 34 sites. This value corresponds to 17 nm when we assume the lattice constant of $a$=5 \AA, which reproduces the experimentally observed $\lambda_{\rm m}$ in MnSi~\cite{Muhlbauer2009}. We also assume that each site, which occupies the unit cell volume of $a^3$, has a spin whose norm is $|S_i|$. The magnitude of magnetization per site is $M_i=\gamma|S_i|$, where $\gamma(\equiv g\mu_{\rm B}/\hbar)$ and $\hbar$ are the gyromagnetic ratio and the Planck constant, respectively. In this study, we adopt $J$=1 meV and $|S_i|/\hbar$=1. The unit conversions when $J$=1 meV and $|S_i|/\hbar$=1 are given in Table I.

Note that this lattice spin model can be derived from a two-dimensional continuum spin model~\cite{Bak1980}, whose energy density is given by,
\begin{align}
\mathcal{E}
&=\mathcal{A}(\nabla \bm m)^2 - M_{\rm s}(\bm m \cdot \bm B_{\rm ext})
\nonumber \\
&+\mathcal{D} \left[
 \left(m_y \frac{\partial m_z}{\partial x} - m_z \frac{\partial m_y}{\partial x}\right)
+\left(m_z \frac{\partial m_x}{\partial y} - m_x \frac{\partial m_z}{\partial y}\right)
\right],
\label{eq:model}
\end{align}
where $\bm m(\bm r)$ is the normalized magnetization vector at spatial position $\bm r$. The first and second terms describe contributions from the ferromagnetic exchange interactions and the Zeeman interactions associated with the external magnetic field $\bm B_{\rm ext}=(0,0,B)$. The third term describes a contribtution from the DM interations. This continuum model is reduced to the lattice spin model by dividing the continuum space into cubic cells with a lattice constant of $a$. The parameters in this continuum spin model is related with those in the above lattice spin model as $J=2a\mathcal{A}$, $D=\mathcal{D}a^2$, and $\bar{\bm B}_{\rm ext}=\bm B_{\rm ext}M_{\rm s}a^3$. The saturation magnetization $M_{\rm s}$ is related with $M$ and $|S_i|$ as $M_{\rm s}=M/a^3=\gamma |S_i|/a^3$.

\begin{figure}[tb]
\includegraphics[scale=1.0]{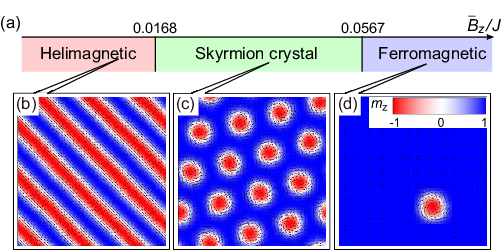}
\caption{(a) Theoretical phase diagram of the spin model in Eq.~(\ref{eq:Hamilt}) with $D/J=0.27$ at $T$=0 as a function of the magnetic field $\bar{B}_z/J$. (b)-(d) Magnetization configurations of (b) helimagnetic, (c) skyrmion crystal, and (d) field-polarized ferromagnetic phases. An isolated skyrmion as a topological defect in the ferromagnetic state is shown in (d).}
\label{Fig03}
\end{figure}
Ground-state phase diagrams of the lattice spin model in Eq.~(\ref{eq:Hamilt}) have been studied as a function of external magnetic field $\bar{B}_z/J$~\cite{SekiBook2016,Iwasaki2013a,Ikka2018}. In the phase diagram for $D/J$=0.27 presented in Fig.~\ref{Fig03}(a), we observe successive emergence of three magnetic phases, i.e., helimagnetic phase, skyrmion crystal phase, and field-polarized ferromagnetic phase with increasing magnetic field $\bar{B}_z$ [see also Figs.~\ref{Fig03}(b)-(d)]. In addition to the crystallized form in the skyrmion crystal phase, skyrmions can emerge as isolated topological defects in the ferromagnetic phase when the system is located in the ferromagnetic phase near the phase boundary to the skyrmion crystal phase [Fig.~\ref{Fig03}(d)]. It is important to note that two threshold magnetic fields of $\bar{B}_{\rm c1}/J$=0.0168 and $\bar{B}_{\rm c2}/J$=0.0567 in Fig.~\ref{Fig03}(a) correspond to $B_{\rm c1}$=147 mT and $B_{\rm c2}$=490 mT in real units, respectively, which reproduce those for a 50-nm thin-plate sample of MnSi observed in experiment~\cite{Tonomura2012}.

We simulate magnetization dynamics induced by spin-orbit torque, exerted by a vertical spin current at $T$=0 using the Landau-Lifshitz-Gilbert-Slonczewski (LLGS) equation,
\begin{align}
\frac{d\bm m_i}{dt}
&=\displaystyle -\gamma \bm m_i \times \bm B_i^{\rm eff} 
+\alpha_{\rm G} \bm m_i \times \frac{d\bm m_i}{dt} 
\nonumber \\ 
&+\frac{\gamma\hbar |\theta_{\rm SH}|j_{\rm e}(\bm r_i)}{2eM_{\rm s}d}
\left[ \bm m_i \times (\bm \sigma_i \times \bm m_i) \right].
\label{eq:LLGS}
\end{align}
The first term is the gyrotropic term, which describes the precessional motion of the magnetization $\bm m_i$ around the effective local magnetic field $\bm B_i^{\rm eff}$. The second term represents Gilbert damping, with a dimensionless damping coefficient fixed at $\alpha_{\rm G}$=0.04 throughout this study. The third term accounts for the spin-orbit torque arising from the vertical spin current with local spin polarization $\bm \sigma_i$. Here $d$ is the thickness of the chiral-ferromagnet layer, and we assume $d$=$a$ in this study. Although the spin-transfer torque also affects the magnetization dynamics, its effect is much weaker than that of the spin-orbit torque in the multilayer system. Therefore, we neglect the contribution of the spin-transfer torque in this study. Here $j_{\rm e}(\bm r_i)$ is the local electric current density, and $e(>0)$ is the elementary charge. 

The vertical spin current is generated from the electric current injected to the heavy-metal layer via the spin Hall effect. The efficiency of conversion from the electric current to the spin current is represented by the spin Hall angle $|\theta_{\rm SH}|=j_{\rm s}/j_{\rm e}$ where $j_{\rm e}$ and $j_{\rm s}$ are their densities. We set $\theta_{\rm SH}$=+0.1 throughout this work. We assume that the system has a spin $S_i$ whose norm is $|S_i|/\hbar=1$ in the $i$th unit cube of $a^3$, which corresponds to the saturation magnetization of $M_{\rm s}=\gamma \hbar/a^3$, where $\gamma$ and $\hbar$ are the gyromagnetic ratio and Planck constant, respectively. The effective local magnetic field $\bm B_i^{\rm eff}$ acting on the magnetization $\bm m_i$ is calculated from the Hamiltonian as,
\begin{align}
\bm B_i^{\rm eff}
=-\frac{1}{M_{\rm s}a^3}\frac{\partial \mathcal{H}}{\partial \bm m_i}
=-\frac{1}{\gamma \hbar}\frac{\partial \mathcal{H}}{\partial \bm m_i}.
\end{align}

The local spin polarization $\bm \sigma_i$ of the vertical spin current should be within the plane and perpendicular to the local current-density vector $\bm j_{\rm e}(\bm r_i)$ in the heavy-metal layer, which is thus given by,
\begin{align}
\bm \sigma_i={\rm sgn}(\theta_{\rm SH}) \frac{\bm j_{\rm e}(\bm r_i) \times \hat{\bm z}}{|\bm j_{\rm e}(\bm r_i) \times \hat{\bm z}|}.
\end{align}
We assume that the current-density vector $\bm j_{\rm e}(\bm r)$ is proportional to the local electric field $\bm E(\bm r)=-\nabla \phi(\bm r)$ with $\phi(\bm r)$ being the scholar electric potential, which leads to $\bm j_{\rm e}(\bm r) \propto \nabla \phi(\bm r)$. For a steady current distribution, the conservation law of current, i.e., $\bm \nabla \cdot \bm j_{\rm e}=0$, holds, which eventually leads to the Poisson equation $\Delta\phi(\bm r)$=0.

We solve this equation using the finite-element method for a nanotrack-shaped system by imposing the following boundary conditions,
\begin{align}
&\phi=
\begin{cases}
 0        & (\text{at the left end of sample})\\
\phi_0 & (\text{at the right end of sample})
\end{cases}
\nonumber \\
&\frac{\partial \phi}{\partial \bm n}=0  \quad\quad (\text{otherwise}).
\label{eq:BC}
\end{align}
The first two conditions imply that the electric current flows from right to left in the nanotrack, corresponding to electron flow from left to right. The third condition reflects the law of charge conservation. Here, the differentiation $\partial \phi/\partial \bm n$ denotes the spatial derivative of the electric potential $\phi$ at the sample boundary, taken along the direction normal to the boundary. We employ a nanotrack system of $1000 \times 50 \times 1$ sites with a triangular notch to calculate the spatial profile of the electric current $\bm j_{\rm e}(\rm r_i)$, while focusing on the region of $600 \times 50 \times 1$ sites around the notch to simulate the current-induced magnetization dynamics. The magnitude of the electric current flowing through the system is specified by the current density $j_{\rm e}$ at either the right or left end of the nanotrack.

In order to count the number of created skyrmions, we calculate the time evolution of the total topological charge $N_{\rm sk}$, give by,
\begin{align}
N_{\rm sk}&=\frac{1}{4\pi}
\displaystyle  \sum_{i=1}^N \bm m_i \cdot
(\bm m_{i+\hat{x}} \times \bm m_{i+\hat{x}+\hat{y}}),
\label{eq:TopoC}
\end{align}
where $N$ is the total number of sites. Even a steady system in equilibrium prior to skyrmion creation exhibits a nonzero topological charge due to the winding alignment of magnetization along the system edges in the presence of DM interactions. We define the bulk topological charge $N_{\rm sk}$ as the value obtained after subtracting this edge contribution.

\section{Results and Discussion}
\begin{figure}[tb]
\includegraphics[scale=0.5]{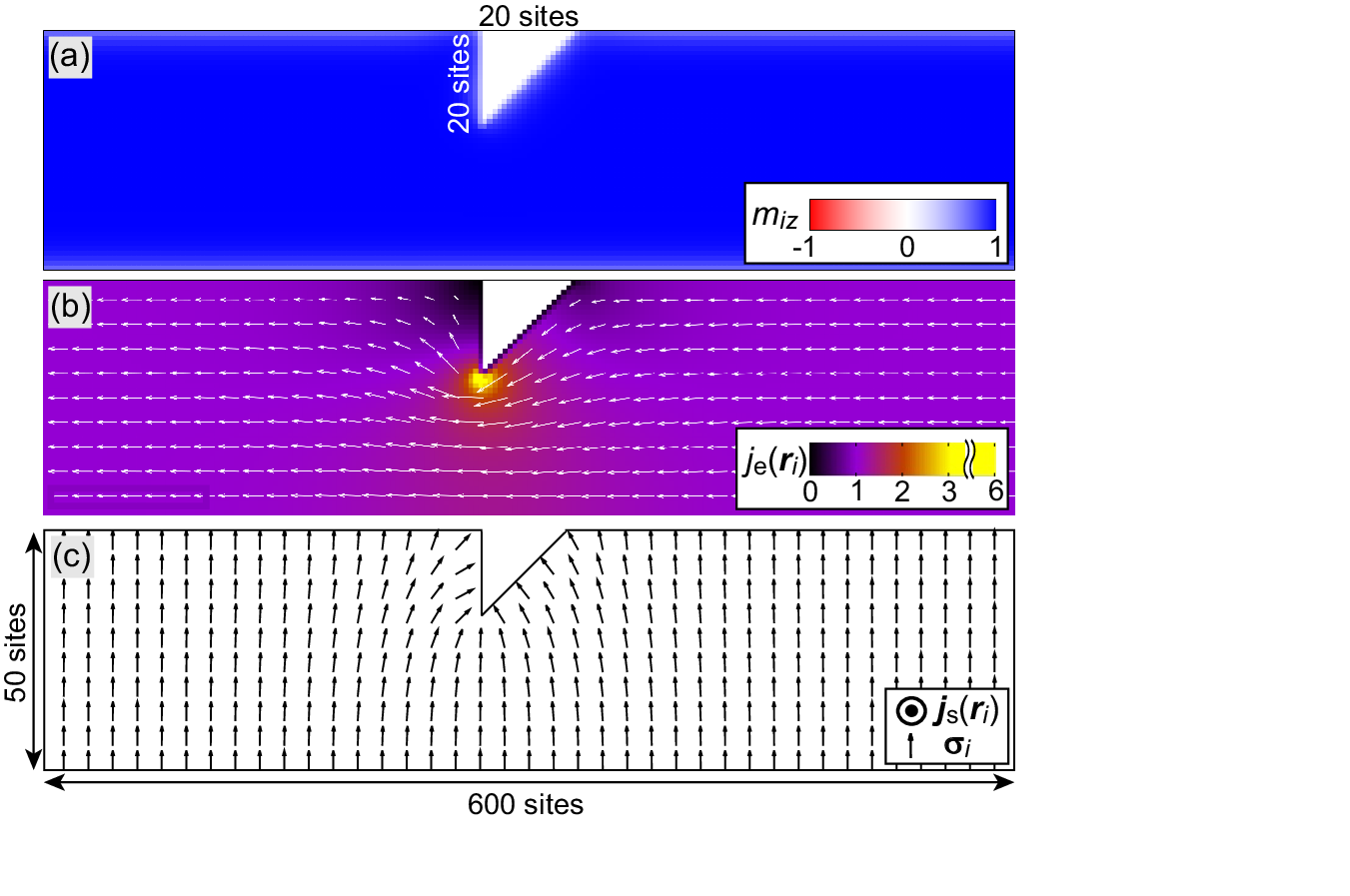}
\caption{(a) Chiral-ferromagnet layer of the nanostrip-shaped magnetic bilayer heterojunction system with a triangular notch. (b) Spatial distribution of the electric current $\bm j_{\rm e}(\bm r_i)$ calculated using the Poisson equation. The local density $j_{\rm e}(\bm r_i)$ is represented by color, while the local current direction $\bm j_{\rm e}(\bm r_i)/j_{\rm e}(\bm r_i)$ is indicated by arrows. (c) Spatial distribution of the local spin polarization $\bm \sigma_i$ of the vertical spin current $\bm j_{\rm s}(\bm r_i)$ indicated by arrows. The local spin current density $j_{\rm s}(\bm r_i)$ is assumed to be proportional to $j_{\rm e}(\bm r_i)$.}
\label{Fig04}
\end{figure}
Figure~\ref{Fig04}(a) shows the system used for the micromagnetic simulations in this study. It consists of a chiral-ferromagnet layer in a forced ferromagnetic state, formed in a nanotrack-shaped magnetic bilayer heterojunction with a triangular notch. By solving the Poisson equation for this system, the spatial distribution of the electric current $\bm j_{\rm e}(\bm r_i)$ in the underlying heavy-metal layer is obtained, as shown in Fig.~\ref{Fig04}(b). Here, the amplitude of the local current density $j_{\rm e}(\bm r_i)$ is represented by color, while the direction of the local current $\bm j_{\rm e}(\bm r_i)/j_{\rm e}(\bm r_i)$ is indicated by arrows. Subsequently, we obtain the spatial distribution of the vertical spin current $\bm j_{\rm s}(\bm r_i)$, as shown in Fig.~\ref{Fig04}(c), where the local spin polarization $\bm \sigma(\bm r_i)$, perpendicular to the electric current $\bm j_{\rm e}(\bm r_i)$, is indicated by arrows. The local spin current density $j_{\rm s}(\bm r_i)$ is assumed to be proportional to the local electric current density $j_{\rm e}(\bm r_i)$ as $j_{\rm s}(\bm r_i)=\theta_{\rm SH}j_{\rm e}(\bm r_i)$.

The vertical spin current with in-plane spin polarization gives rise to a magnetic torque, i.e., the spin-orbit torque, acting on magnetizations in the chiral-magnet layer, which causes magnetization reversal. As seen in Fig.~\ref{Fig04}(b), the electric current density $j_{\rm e}(\rm r_i)$ and eventually the spin current density $j_{\rm s}(\rm r_i)$ are significantly enhanced at the corner of the notch. This indicates that the notch works to locally enhance the spin current density. In addition, the notch plays a role to relax the topological constraints by introducing discontinuity in the spatial distribution of magnetizations. Concequently, the magnetizations around the corner of the notch becomes capable of rotating gradually and being reversed locally, which result in the formation of skyrmion core. Once a skyrmion core is created, magnetizations around the core become rotating spontanesouly because of the DM interactions and form a particle-like skyrmion configuration in the presence of external magnetic field.

The vertical spin current with in-plane spin polarization exerts a magnetic torque, i.e., the spin-orbit torque, on the magnetization in the chiral-ferromagnet layer, resulting in magnetization reversal. As seen in Fig.~\ref{Fig04}(b), the electric current density $j_{\rm e}(\rm r_i)$, and consequently the spin current density $j_{\rm s}(\rm r_i)$, is significantly enhanced at the corner of the notch. This indicates that the notch serves to locally increase the spin current density. In addition, the notch relaxes topological constraints by introducing a discontinuity in the spatial distribution of the magnetization. As a result, the magnetization near the notch corner can gradually rotate and reverse locally, leading to the formation of a skyrmion core. Once a skyrmion core is created, the surrounding magnetization begins to rotate spontaneously due to DM interactions, resulting in the formation of a particle-like skyrmion configuration in the presence of an external magnetic field.

\begin{figure*}[tb]
\includegraphics[scale=1.0]{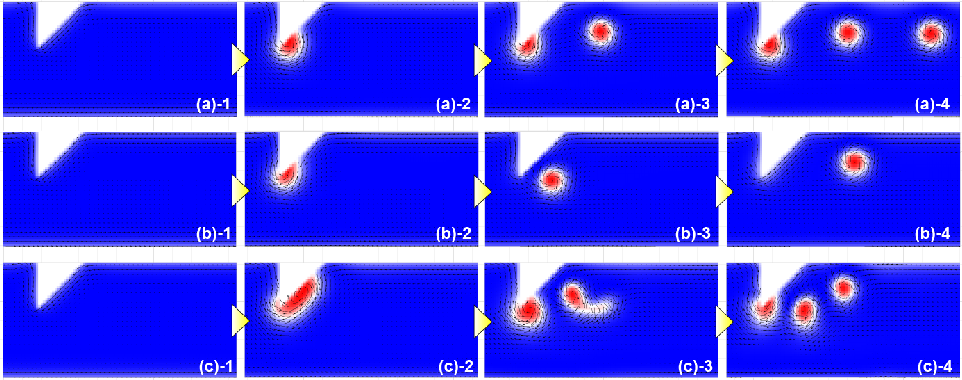}
\caption{Snapshots at selected moments for various skyrmion creation processes with different spin current densities $j_{\rm s}(\equiv \theta_{\rm SH}j_{\rm e})$. Here, the electric current is continuously injected, and the magnetic field is fixed at $\bar{B}_z/J=0.06$. (a) Successive skyrmion creation for $j_{\rm s}=1.2 \times 10^{10}$ A/m$^2$. (b) Single skyrmion creation for $j_{\rm s}=1.1 \times 10^{10}$ A/m$^2$. The skyrmion creation occurs only once, despite the continuous presence of the spin current. (c) Succesive skyrmion creation for $j_{\rm s}=2.1 \times 10^{10}$ A/m$^2$. When $j_{\rm e}$ exceeds a certain threshold, the creation rate decreases because a newly formed skyrmion seed at the notch interferes with the previously created skyrmion.}
\label{Fig05}
\end{figure*}
In Fig.~\ref{Fig05}, we present snapshots at selected moments of various skyrmion creation processes under different spin current densities $j_{\rm s}(\equiv \theta_{\rm SH}j_{\rm e})$. In the simulations, we assume a continuously injected electric current, resulting in a steady spin current. The external magnetic field is fixed at $\bar{B}_z/J=0.06$. Figure~\ref{Fig05}(a) shows the case of successive skyrmion creation for  $j_{\rm s}=1.2 \times 10^{10}$ A/m$^2$. When the electric current density is sufficiently large, local magnetization reversal occurs repeatedly at the corner of the triangular notch. Once a skyrmion is formed, it moves rightward along the longitudinal direction of the track, away from the notch, thereby allowing the next skyrmion to be created.

Figure~\ref{Fig05}(b) shows a single skyrmion creation when $j_{\rm s}$ is slightly smaller as $j_{\rm s}=1.1 \times 10^{10}$ A/m$^2$. By tuning the current density $j_{\rm e}$, we can achieve a condition in which  only one skyrmion is created, despite the presence of a steady spin current. In this case, the first skyrmion is created because the spin current applied to the initial magnetization configuration, which corresponds to the relaxed state in the absence of the spin current, acts as an impulsive force, triggering local magnetization reversal at the notch. However, after the first skyrmion is created and moves away from the notch region, the magnetization configuration relaxes into a steady state under the continuous spin current. As a result, the further application of the spin current can no longer induce additional magnetization reversal, leading to the observed behavior in which no further skyrmions are created.

Applying a larger electric current does not always lead to faster successive skyrmion creation. In fact, the creation rate, that is, the number of skyrmions created per unit time, decreases when the electric current density exceeds a certain threshold value. In the presence of a large electric current, local magnetization reversal occurs rapidly in succession. However, if magnetization reversal takes place immediately after a skyrmion has been created and before it has moved sufficiently far from the notch, the expansion of the region with reversed magnetization, which would grow into the next skyrmion seed, is suppressed by the repulsive potential force from the existing skyrmion. This repulsive interaction interferes with the successive creation process and thus reduce the creation rate. Figure~\ref{Fig05}(c) illustrates such a case for $j_{\rm s}=2.1 \times 10^{10}$ A/m$^2$, which exceeds the threshold value.

\begin{figure}[tb]
\includegraphics[scale=1.0]{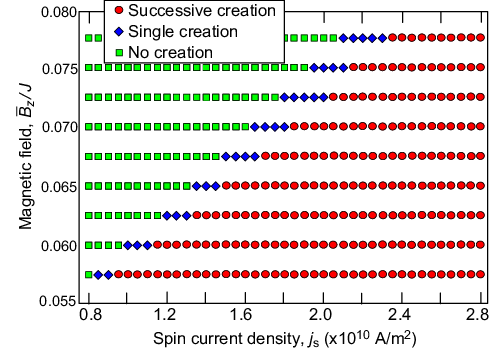}
\caption{Skyrmion-creation diagram in the plane of the external magnetic field $\bar{B}_z$ and the spin current density $j_{\rm s}(\equiv \theta_{\rm SH}j_{\rm e})$. Under continuous electric current injection, three distinct types of behavior are observed.}
\label{Fig06}
\end{figure}
Figure~\ref{Fig06} shows a diagram of the types of skyrmion creation in plane of the electric current density $j_{\rm e}$ and the external magnetic field $\bar{B}_z$ applied perpendicular to the plane. We observe three distinct types of behavior under continuous electric current injection. In the region with red circles, successive skyrmion creation occurs, whereas in the region with green squares, no creation is observed. Between these two regions, a single skyrmion is created shortly after the electric current is applied, but no further skyrmions are created afterwards. We also observe that the threshold current density depends on the strength of the external magnetic field $\bar{B}_z$. A larger electric current is required for skyrmion creation under a stronger magnetic field as a stronger torque is necessary to induce magnetization reversal.

\begin{figure}[tb]
\includegraphics[scale=1.0]{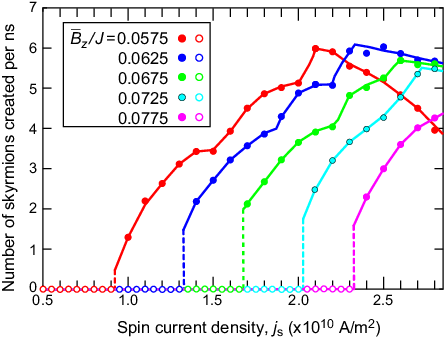}
\caption{Skyrmion creation rates, defined as the number of skyrmions created per nanosecond, plotted as a function of the electric current density $j_{\rm e}$ for various strengths of the external magnetic field $\bar{B}_z$. Solid lines are provided as guides to the eyes.}
\label{Fig07}
\end{figure}
In Fig.~\ref{Fig07}, we plot the calculated rates of skyrmion creations, i.e., the numbers of created skyrmions per nanosecond as functions of the spin current density $j_{\rm s}(\equiv \theta_{\rm SH}j_{\rm e})$ for various strengths of the external magnetic field $\bar{B}_z$. Here the creation rate in the case of single creation is set to be zero because only one skyrmion can be created  no matter how long the current is applied. We realize several features in this figure. First, as mentioned above, there appears a threshold current density $j_{\rm s,c1}$ for the skyrmion creation. The creation rate is zero below $j_{\rm s,c1}$, while it exhibits an abrupt change or even a jump at $j_{\rm s,c1}$. Moreover, the value of $j_{\rm s,c1}$ is larger as the magnetic field $\bar{B}_z$ is stronger. Second, the creation rate increases as the current density $j_{\rm s}$ increases as long as $j_{\rm s}$ is smaller than another threshold value $j_{\rm s,c2}$. Third, when the current density exceeds the threshold $j_{\rm s,c2}$, the creation rate starts decreasing with further increasing $j_{\rm s}$. The value of $j_{\rm s,c2}$ also increases as $\bar{B}_z$ increases.

In Fig.~\ref{Fig07}, we plot the calculated skyrmion creation rates, defined as the number of skyrmions created per nanosecond, as a function of the spin current density $j_{\rm s}(\equiv \theta_{\rm SH}j_{\rm e})$ for various strengths of the external magnetic field $\bar{B}_z$. The creation rate for the case of a single skyrmion creation is set to zero, as only one skyrmion can be created regardless of how long the current is applied. Several features are observed in this figure. First, as mentioned above, a threshold current density $j_{\rm s,c1}$ for skyrmion creation is evident. The creation rate is zero below $j_{\rm s,c1}$, while it shows an abrupt change or even a jump at $j_{\rm s,c1}$. Moreover, the value of $j_{\rm s,c1}$ increases as the magnetic field $\bar{B}_z$ increases. Second, the creation rate increases as the current density $j_{\rm s}$ increases as long as $j_{\rm s}$ is smaller than another threshold value $j_{\rm s,c2}$. Third, when the current density exceeds $j_{\rm s,c2}$, the creation rate begins to decrease with further increasing $j_{\rm s}$. The value of $j_{\rm s,c2}$ also increases as $\bar{B}_z$ increases.

\begin{figure}[tb]
\includegraphics[scale=1.0]{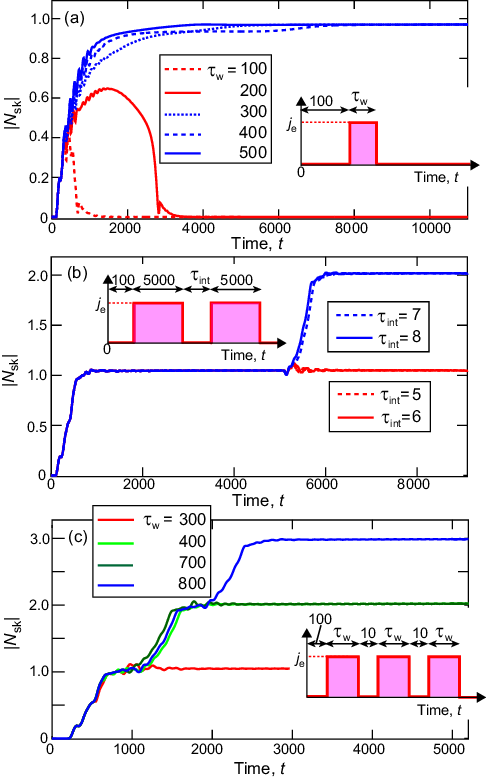}
\caption{(a) Time profiles of the topological charge $|N_{\rm sk}|$ after the application of a single current pulse for various pulse durations $\tau_{\rm w}$. The pulse duration $\tau_{\rm w}$ is varied from 100 to 500 in increments of 100. (b) Time profiles of $|N_{\rm sk}|$ when two current pulses are applied successively with varying intervals $\tau_{\rm int}$. The interval $\tau_{\rm int}$ between the two pulses is varied from 5 to 8 in increments of 1. (c) Time profiles of $|N_{\rm sk}|$ when multiple current pulses are applied sequentially. The pulse duration $\tau_{\rm w}$ is varied from 300 to 800 in increments of 100. In all simulations, the amplitude of the spin-current pulse and the external magnetic field are fixed at $j_{\rm s}=1.1 \times 10^{10}$ A/m$^2$ and $\bar{B}_z/J$=0.06, respectively.}
\label{Fig08}
\end{figure}
Finally, we examine various types of current pulse applications, i.e., single-pulse application, twin -pulse applications, and sequential pulse applications. Figures~\ref{Fig08}(a)-(c) show simulated time profiles of the topological number $|N_{\rm sk}|$ for respective cases. For these simulations, the external magnetic field is fixed at $\bar{B}_z/J$=0.06, and the amplitude of current pulse is fixed at $j_{\rm s}=1.1 \times 10^{10}$ A/m$^2$. This current density corresponds to that for the single skyrmion creation for $\bar{B}_z/J$=0.06. Namely, only one skyrmion can be created if we inject an electric current of this current density in a continuous manner. By the simulations for Fig.~\ref{Fig08}(a), we investigate the minimum duration $\tau_{\rm w}^{\rm min}$ of the single pulse required for creating a skyrmion by varying the pulse duration from $\tau_{\rm w}$=100 to 500 in increments of 100. In Fig.~\ref{Fig08}(a), we find that $\tau_{\rm w}^{\rm min}$ takes a value between 200 and 300, which correspond to 0.132 ns and 0.198 ns, respectively.

Finally, we examine various types of current pulse applications: single-pulse, twin-pulse, and sequential-pulse applications. Figures~\ref{Fig08}(a)-(c) show the simulated time profiles of the topological charge $|N_{\rm sk}|$ for each case. In these simulations, the external magnetic field is fixed at $\bar{B}_z/J$=0.06, and the amplitude of the current pulse is fixed at $j_{\rm s}=1.1 \times 10^{10}$ A/m$^2$. This current density corresponds to the condition under which only a single skyrmion is created when the current is applied continuously at $\bar{B}_z/J$=0.06. In the simulations for Fig.~\ref{Fig08}(a), we investigate the minimum pulse duration $\tau_{\rm w}^{\rm min}$ required for skyrmion creation by varying the pulse duration $\tau_{\rm w}$ from 100 to 500 in increments of 100. The results indicate that $\tau_{\rm w}^{\rm min}$ lies between 200 and 300, which correspond to 0.132 ns and 0.198 ns, respectively.

In the simulations for Fig.\ref{Fig08}(b), we investigate the minimum relaxation time required for the creation of a second skyrmion through application of twin current pulses. Specifically, we examine the minimum interval $\tau_{\rm int}^{\rm min}$ between the two pulses for successfully creating another skyrmion by the second pulse by varying the interval $\tau_{\rm int}$ in increments of 1. The results shown in Fig.~\ref{Fig08}(b) indicate that $\tau_{\rm int}^{\rm min}$ ies between 6 and 7, which correspon to 3.96 ps and 4.62 ps, respectively.

In the simulations for Fig.~\ref{Fig08}(c), we examine the effect of sequential current pulse applications with a fixed interval of $\tau_{\rm int}$=10. The pulse duration $\tau_{\rm w}$ is varied from 300 to 800 in increments of 100. Note that this interval $\tau_{\rm int}$=10 is longer than the minimum interval $\tau_{\rm int}^{\rm min}$ required for creating both the first and second skyrmions by twin-pulse application. Similarly, the pulse durations in the range $300 \le \tau_{\rm w} \le 800$ all exceed the minimum pulse duration $\tau_{\rm w}^{\rm min}$ necessary for skyrmion creation by single-pulse application. However, these two conditions do not necessarily guarantee that skyrmions will be created successively with a constant rate. In fact, as shown in Fig.~\ref{Fig08}(c), the skyrmion creation exhibits complex behavior in response to the triple-pulse application. For relatively short pulses with $\tau_{\rm w}$=300, only the first pulse creates a skyrmion, while the subsequent two do not. For moderately long pulses of $\tau_{\rm w}$=400-700, the first and second pulses each create a skyrmion, but the third fails. This is because if the pulse duration is not sufficiently long, the previously created skyrmion may still be located near the notch when the next skyrmion seed begins to form. As discussed above, in such cases, the repulsive potential from the existing skyrmions hinders the growth of the new seed into a complete skyrmion. Moreover, if two skyrmions are created too close to one another, their mutual repulsive interactions cause orbital motion around each other. This rotational motion inhibits their smooth departure from the notch area and further complicates the skyrmion creation process. We find that, under the present conditions, a pulse duration of at least $\tau_{\rm w} \ge 800$ is required to achieve one-to-one skyrmion creation corresponding to the number of applied pulses. If the system parameters or external conditions change, the pulse duration must be accordingly adjusted to ensure consistent and successive skyrmion creation with sequential current pulses.

\section{Conclusion}
In conclusion, we have theoretically proposed a practically viable method for the controlled creation of magnetic skyrmions through electric current injection utilizing spin-orbit torque. In the considered magnetic heterojunction composed of a chiral-ferromagnet layer and a nonmagnetic heavy-metal layer, the in-plane electric current injected into the heavy-metal layer is converted into a vertical spin current via the spin Hall effect, owing to the strong spin-orbit coupling in the heavy-metal layer. The resulting spin current exerts spin-orbit torque on the magnetizations in the chiral-ferromagnet layer, enabling the creation and manipulation of magnetic skyrmions. When a conventional ferromagnet is used as the magnetic layer, N\'{e}el-type skyrmions are expected to be stabilized due to the interfacial DM interaction. However, such systems are not suitable for skyrmion creation via spin-orbit torque, because the created N\'{e}el-type skyrmions tend to move perpendicular to the electric curren, i.e., across the width direction of the nanotrack rather than along its length direction. As a result, skyrmions created near a notch move toward the opposite longitudinal edge, where they would be absorbed and vanish upon collision. To solve this problem, we have proposed an alternative magnetic heterojunction system with the magnetic layer composed of a chiral ferromagnet such as MnSi, Fe$_{1-x}$Co$_x$Si, or FeGe. In this configuration, Bloch-type skyrmions are stabilized by the bulk DM interactions intrinsic to the chiral crystal structure. Unlike N\'{e}el-type skyrmions, Bloch-type skyrmions driven by spin-orbit torque move parallel to the electric current, i.e., along the nanotrack, which allows them to persist after creation. These robustly propagating skyrmions can therefore be harnessed as information carriers in skyrmion-based racetrack memory devices.

\section{Acknowledgment}
MM thanks Shiho Nakamura, Xichao Zhang, and Xiuzhen Yu for fruitful discussions. We would also like to thank Xichao Zhang for his assistance in preparing Fig. 1 and Fig. 2.
This work was supported by 
the cooperation of organization between Kioxia Corporation and Waseda University, 
JSPS KAKENHI (No.~20H00337, No.~24H02231 and No.~25H00611),  
JST CREST (No.~JPMJCR20T1), and 
Waseda University Grant for Special Research Projects (No.~2022C-139 and No.~2025C-133).

\section{Author Contributions}
M.M. conceptualized and supervised the work. M.M. developed the software. Y.U. conducted the simulations and analyzed the data. Y.U. and M.M. made the figures. M.M. wrote the manuscript.

\end{document}